\documentclass[longauth]{aa} 
\usepackage{graphicx,natbib}
\usepackage{txfonts}
\begin{document}
   \title{A new activity phase of the blazar \object{3C 454.3}}

   \subtitle{Multifrequency observations by the WEBT and  XMM-Newton in 2007--2008\thanks{The radio-to-optical 
   data presented in this paper are stored in the WEBT archive; for questions regarding their availability,
   please contact the WEBT President Massimo Villata ({\tt villata@oato.inaf.it}).}}

   \author{C.~M.~Raiteri              \inst{ 1}
   \and   M.~Villata                  \inst{ 1}
   \and   V.~M.~Larionov              \inst{ 2,3}
   \and   M.~A.~Gurwell               \inst{ 4}
   \and   W.~P.~Chen                  \inst{ 5}
   \and   O.~M.~Kurtanidze            \inst{ 6}
   \and   M.~F.~Aller                 \inst{ 7}
   \and   M.~B\"ottcher               \inst{ 8}
   \and   P.~Calcidese                \inst{ 9}
   \and   F.~Hroch                    \inst{10}
   \and   A.~L\"ahteenm\"aki          \inst{11}
   \and   C.-U.~Lee                   \inst{12}
   \and   K.~Nilsson                  \inst{13}
   \and   J.~Ohlert                   \inst{14}
   \and   I.~E.~Papadakis             \inst{15,16}
   \and   I.~Agudo                    \inst{17}
   \and   H~.D.~Aller                 \inst{ 7}
   \and   E.~Angelakis                \inst{18}
   \and   A.~A.~Arkharov              \inst{ 3}
   \and   U.~Bach                     \inst{18}
   \and   R.~Bachev                   \inst{19}
   \and   A.~Berdyugin                \inst{13}
   \and   C.~S.~Buemi                 \inst{20}
   \and   D.~Carosati                 \inst{21}
   \and   P.~Charlot                  \inst{22,23}
   \and   E.~Chatzopoulos             \inst{16}
   \and   E.~Forn\'e                  \inst{24}
   \and   A.~Frasca                   \inst{20}
   \and   L.~Fuhrmann                 \inst{18}
   \and   J.~L.~G\'{o}mez             \inst{17}
   \and   A.~C.~Gupta                 \inst{25}
   \and   V.~A.~Hagen-Thorn           \inst{ 2}
   \and   W.-S. Hsiao                 \inst{ 5}
   \and   B.~Jordan                   \inst{26}
   \and   S.~G.~Jorstad               \inst{27}
   \and   T.~S.~Konstantinova         \inst{ 2}
   \and   E.~N.~Kopatskaya            \inst{ 2}
   \and   T.~P.~Krichbaum             \inst{18}
   \and   L.~Lanteri                  \inst{ 1}
   \and   L.~V.~Larionova             \inst{ 2}
   \and   G.~Latev                    \inst{28}
   \and   J.-F.~Le~Campion            \inst{22,23}
   \and   P.~Leto                     \inst{29}
   \and   H.-C.~Lin                   \inst{ 5}
   \and   N.~Marchili                 \inst{18}
   \and   E.~Marilli                  \inst{20}
   \and   A.~P.~Marscher              \inst{27}
   \and   B.~McBreen                  \inst{30}
   \and   B.~Mihov                    \inst{19}
   \and   R.~Nesci                    \inst{31}
   \and   F.~Nicastro                 \inst{32}
   \and   M.~G.~Nikolashvili          \inst{ 6}
   \and   R.~Novak                    \inst{33}
   \and   E.~Ovcharov                 \inst{28}
   \and   E.~Pian                     \inst{34}
   \and   D.~Principe                 \inst{ 8}
   \and   T.~Pursimo                  \inst{35}
   \and   B.~Ragozzine                \inst{ 8}
   \and   J.~A.~Ros                   \inst{24}
   \and   A.~C.~Sadun                 \inst{36}
   \and   R.~Sagar                    \inst{25}
   \and   E.~Semkov                   \inst{19}
   \and   R.~L.~Smart                 \inst{ 1}
   \and   N.~Smith                    \inst{37}
   \and   A.~Strigachev               \inst{19}
   \and   L.~O.~Takalo                \inst{13}
   \and   M.~Tavani                   \inst{38}
   \and   M.~Tornikoski               \inst{11}
   \and   C.~Trigilio                 \inst{20}
   \and   K.~Uckert                   \inst{ 8}
   \and   G.~Umana                    \inst{20}
   \and   A.~Valcheva                 \inst{19}
   \and   S.~Vercellone               \inst{39}
   \and   A.~Volvach                  \inst{40}
   \and   H.~Wiesemeyer               \inst{41}
 }

   \offprints{C.~M.~Raiteri}

   \institute{
          INAF, Osservatorio Astronomico di Torino, Italy                                                     
   \and   Astron.\ Inst., St.-Petersburg State Univ., Russia                                                  
   \and   Pulkovo Observatory, St.\ Petersburg, Russia                                                        
   \and   Harvard-Smithsonian Center for Astroph., Cambridge, MA, USA                                         
   \and   Institute of Astronomy, National Central University, Taiwan                                         
   \and   Abastumani Astrophysical Observatory, Georgia                                                       
   \and   Department of Astronomy, University of Michigan, MI, USA                                            
   \and   Department of Physics and Astronomy, Ohio Univ., OH, USA                                            
   \and   Osservatorio Astronomico della Regione Autonoma Valle d'Aosta, Italy                                
   \and   Inst.\ of Theor.\ Phys.\ and Astroph., Masaryk Univ., Czech Republic                                
   \and   Mets\"ahovi Radio Obs., Helsinki Univ.\ of Technology, Finland                                      
   \and   Korea Astronomy and Space Science Institute, South Korea                                            
   \and   Tuorla Observatory, Univ.\ of Turku, Piikki\"{o}, Finland                                           
   \and   Michael Adrian Observatory, Trebur, Germany                                                         
   \and   IESL, FORTH, Heraklion, Crete, Greece                                                               
   \and   Physics Department, University of Crete, Greece                                                     
   \and   Instituto de Astrof\'{i}sica de Andaluc\'{i}a (CSIC), Granada, Spain                                
   \and   Max-Planck-Institut f\"ur Radioastronomie, Bonn, Germany                                            
   \and   Inst.\ of Astronomy, Bulgarian Academy of Sciences, Sofia, Bulgaria                                 
   \and   INAF, Osservatorio Astrofisico di Catania, Italy                                                    
   \and   Armenzano Astronomical Observatory, Italy                                                           
   \and   Universit\'e de Bordeaux, Observatoire Aquitain des Sciences de l'Univers, Floirac, France          
   \and   CNRS, Laboratoire d'Astrophysique de Bordeaux -- UMR 5804, Floirac, France                          
   \and   Agrupaci\'o Astron\`omica de Sabadell, Spain                                                        
   \and   ARIES, Manora Peak, Nainital, India                                                                 
   \and   School of Cosmic Physics, Dublin Institute For Advanced Studies, Ireland                            
   \and   Institute for Astrophysical Research, Boston University, MA, USA                                    
   \and   Sofia University, Bulgaria                                                                          
   \and   INAF, Istituto di Radioastronomia, Sezione di Noto, Italy                                           
   \and   School of Physics, University College Dublin, Ireland                                               
   \and   Dept.\ of Phys.\ ``La Sapienza" Univ, Roma, Italy                                                   
   \and   INAF, Osservatorio Astronomico di Roma, Italy                                                       
   \and   N.\ Copernicus Observatory and Planetarium in Brno, Czech Republic                                  
   \and   INAF, Osservatorio Astronomico di Trieste, Italy                                                    
   \and   Nordic Optical Telescope, Santa Cruz de La Palma, Spain                                             
   \and   Dept.\ of Phys., Univ.\ of Colorado Denver, Denver, CO USA                                          
   \and   Cork Institute of Technology, Cork, Ireland                                                         
   \and   INAF, IASF-Roma, Italy                                                                              
   \and   INAF, IASF-Milano, Italy                                                                            
   \and   Radio Astronomy Lab.\ of Crimean Astrophysical Observatory, Ukraine                                 
   \and   Instituto de Radioastronom\'{i}a Millim\'{e}trica, Granada, Spain                                   
 }

   \date{}
 
  \abstract
{}
   {The Whole Earth Blazar Telescope (WEBT) consortium has been monitoring the blazar \object{3C 454.3} 
from the radio to the optical bands since 2004 to study its emission variability properties.}
   {We present and analyse the multifrequency results of the 2007--2008 observing season, 
including XMM-Newton observations and near-IR spectroscopic monitoring, and compare the recent emission behaviour with the past one. The historical mm light curve is presented here for the first time.}
   {In the optical band we observed a multi-peak outburst in July--August 2007, 
and other faster events in November 2007 -- February 2008. During these outburst phases, several episodes of intranight variability were detected.
A mm outburst was observed starting from mid 2007, whose rising phase was contemporaneous to the optical brightening. A slower flux increase also affected the higher radio frequencies, the flux enhancement disappearing below 8 GHz. 
The analysis of the optical-radio correlation and time delays, as well as the behaviour of the mm light curve, confirm our previous predictions, suggesting that changes in the jet orientation likely occurred in the last few years. The historical multiwavelength behaviour indicates that a significant variation in the viewing angle may have happened around year 2000.
Colour analysis reveals a complex spectral behaviour, which is due to the interplay of different emission components.
All the near-IR spectra show a prominent H$\alpha$ emission line ($\rm EW_{obs}$ = 50--120 \AA), whose flux appears nearly constant, indicating that the broad line region is not affected by the jet emission.
We show the broad-band SEDs corresponding to the epochs of the XMM-Newton pointings and compare them to those obtained at other epochs, when the source was in different brightness states. 
A double power-law fit to the EPIC spectra including extra absorption suggests that the soft-X-ray spectrum is concave, and that the curvature becomes more pronounced as the flux decreases. This connects fairly well with the UV excess, which becomes more prominent with decreasing flux. The most obvious interpretation implies that, as the beamed synchrotron radiation from the jet dims, we can see both the head and the tail of the big blue bump.
The X-ray flux correlates with the optical flux, suggesting that in the inverse-Compton process either the seed photons are synchrotron photons at IR--optical frequencies or the relativistic electrons are those that produce the optical synchrotron emission. The X-ray radiation would thus be produced in the jet region from where the IR--optical emission comes.
}
  {}



   \maketitle
%

\section{Introduction}

Blazars (i.e.\ flat-spectrum radio quasars and BL Lacertae objects) constitute a special class
of radio-loud active galactic nuclei (AGNs), whose emission is dominated by relativistically-beamed non-thermal radiation from a plasma jet. 
These objects are observed at all wavelengths, from the radio band up to $\gamma$-ray energies, with variability time scales ranging from hours to years. 
The low-energy non-thermal emission, from the radio band to the optical
(sometimes up to UV or X-ray) frequencies, is due to synchrotron radiation, while the higher-energy emission is
likely to be produced by inverse-Compton scattering. 
Shocks travelling down the jet are a possible mechanism that explains flux variability, 
probably coupled with geometrical effects \citep[see e.g.][]{vil02,mar08}.
A UV excess is observed in some cases, which is the signature of the thermal radiation from the accretion disc feeding the central black hole.

The flat-spectrum radio quasar \object{3C 454.3} is one of the most studied blazars, especially after observation of an exceptional bright state from the mm wavelengths to the X-ray frequencies in May 2005.
This outburst was monitored from the radio to the optical bands by the Whole Earth Blazar Telescope (WEBT)\footnote{{\tt http://www.oato.inaf.it/blazars/webt/} \\ see e.g.\ \citet{vil04a,boe07,rai08a}.}, whose observations were published by \citet{vil06}\footnote{Observations by the INTEGRAL and Swift satellites and by the Rapid Eye Mount (REM) telescope were presented in \citet{pia06,gio06,fuh06}.}.
The WEBT monitoring continued to follow the subsequent noticeable radio activity; an interpretation of the optical-radio correlation was presented by \citet{vil07}.
In the 2006--2007 observing season the source remained in a faint state. The low contribution by the synchrotron emission from the jet allowed \citet{rai07b} to recognise in the source spectral energy distributions (SEDs) both a little blue bump due to line emission from the broad line region, and a big blue bump due to thermal emission from the accretion disc.

In July 2007 the source showed a new optical outburst phase, motivating the continuation of the ongoing WEBT campaign and triggering observations by the Astro-rivelatore Gamma a Immagini LEggero (AGILE), which detected the source in its brightest $\gamma$-ray state ever observed \citep{ver08a}. 
This new high-energy bright state was compared to other states of the source by \citet{ghi07}, who
explained the emission variability in terms of changes of the site where the radiation is produced in the jet.
A further significant detection by AGILE was announced in November \citep{che07,puc07}, which in turn led to intensified monitoring by the WEBT.
Some results of the WEBT observations in November--December 2007, during the second AGILE detection period, were presented by \citet{rai08b}. A cross-correlation analysis between the $\gamma$-ray fluxes in November 2007 and the WEBT optical data was performed by \citet{ver08b}; a similar analysis on the December 2007 data will be presented by Donnarumma et al.\ (in preparation).

In this paper we show all the multiwavelength data taken during the whole last WEBT campaign, from May 2007 to February 2008. These include radio-to-optical light curves, optical--UV and X-ray data from two pointings of the XMM-Newton satellite in May, and near-IR spectra taken at Campo Imperatore.
The new data trace the recent behaviour of the source, which is compared to the past one. We also present here for the first time the historical mm light curve, which gives information on the source emission in a frequency range, between the radio and optical bands, that is crucial for understanding the source variability.

\section{Radio-to-optical observations by the WEBT}

The list of observatories participating in the last WEBT campaign on 3C 454.3 is shown in Table \ref{obs}, which also reports the size of the telescope and the bands where observations were carried out.

\begin{table}
\caption{Ground-based observatories participating in this work.}
\label{obs}
\centering
\begin{tabular}{l r c  }
\hline\hline
Observatory    & Tel.\ size    & Bands\\
\hline
\multicolumn{3}{c}{\it Optical}\\
Abastumani, Georgia      &  70 cm         & $R$                \\
ARIES, India             & 104 cm         & $BVRI$             \\
Armenzano, Italy         &  35 cm         & $BRI$              \\
Armenzano, Italy         &  40 cm         & $BVRI$             \\
Belogradchik, Bulgaria   &  60 cm         & $VRI$              \\
Bordeaux, France         &  20 cm         & $V$                \\
Calar Alto, Spain$^a$    & 220 cm         & $R$                \\
Catania, Italy           &  91 cm         & $UBV$              \\
Crimean, Ukraine         &  70 cm         & $BVRI$             \\
Kitt Peak (MDM), USA     & 130 cm         & $UBVRI$            \\
L'Ampolla, Spain         &  36 cm         & $R$                \\
Lulin (LOT), Taiwan      & 100 cm        & $BVRI$             \\
Lulin (SLT), Taiwan      &  40 cm        & $R$                \\
Michael Adrian, Germany  & 120 cm        & $R$                \\
Mt.\ Lemmon, USA         & 100 cm        & $BVRI$             \\
New Mexico Skies, USA    &  30 cm        & $VRI$              \\
N.\ Copernicus, Czech Republic  & 40 cm         & $VR$               \\
Roque (KVA), Spain       &  35 cm        & $R$                \\
Roque (NOT), Spain       & 256 cm        & $UBVRI$            \\
Rozhen, Bulgaria         & 50/70 cm      & $BVR$              \\
Rozhen, Bulgaria         & 200 cm        & $BVRI$             \\
Sabadell, Spain          &  50 cm        & $R$                \\
Skinakas, Greece         & 130 cm        & $BVRI$             \\
Sommers-Bausch, USA      &  61 cm        & $VRI$              \\
St.\ Petersburg, Russia  &  40 cm        & $BVRI$             \\
Teide (BRT), Spain       &  35 cm        & $BVR$              \\
Torino, Italy            & 105 cm        & $BVRI$             \\
Valle d'Aosta, Italy     &  81 cm        & $BVRI$             \\
Vallinfreda, Italy       &  50 cm        & $R$                \\
\hline
\multicolumn{3}{c}{\it Near-infrared}\\
Campo Imperatore, Italy  & 110 cm        & $JHK$          \\
Roque (NOT), Spain       & 256 cm        & $JHK$          \\
\hline
\multicolumn{3}{c}{\it Radio}\\
Crimean (RT-22), Ukraine & 22 m          & 37 GHz          \\
Effelsberg, Germany      & 100 m         &1.4, 2.7, 4.9, 8.4,  \\
                         &               &10.5, 14.6, 23, 42 GHz\\
Mauna Kea (SMA), USA     &$8 \times 6$ m$^b$ & 230, 345 GHz      \\
Medicina, Italy          & 32 m          & 5, 8, 22 GHz        \\
Mets\"ahovi, Finland     & 14 m          & 37 GHz              \\
Noto, Italy              & 32 m          & 43 GHz              \\
Pico Veleta, Spain       & 30 m          & 86, 142, 229 GHz    \\
UMRAO, USA               & 26 m          & 4.8, 8.0, 14.5 GHz   \\
\hline
\multicolumn{3}{l}{$^a$ Calar Alto data were acquired as part of the MAPCAT (Monitoring}\\
\multicolumn{3}{l}{AGN with Polarimetry at the Calar Alto Telescopes) project.}\\
\multicolumn{3}{l}{$^b$ Radio interferometer including 8 dishes of 6 m size.}
\end{tabular}
\end{table}

   \begin{figure}
      \caption{Optical $UBVRI$ and near-IR $JHK$ light curves of 3C 454.3 in the 2007--2008 observing
 season. Maximum and minimum brightness levels are marked with horizontal dotted lines.
Vertical lines indicate the two XMM-Newton pointings of May 23 and 31, 2007.
The yellow strip highlights that part of the $R$-band light curve already presented in \citet{rai08b}.}
         \label{ottico-nir}
   \end{figure}

\subsection{Optical and near-IR data}

Optical and near-IR data were acquired in Johnson-Cousins $UBVRI$ and $JHK$ bands.
Magnitude calibration and light curve construction were performed following \citet{rai08b}.
The result is displayed in Fig.\ \ref{ottico-nir}. 

Over the whole 2007--2008 observing season, the source showed noticeable variability, with a major outburst in July--August 2007, and other faster events in November 2007 -- February 2008. 
The brightest level, $R=12.58$ observed on December 1, 2007, is only 0.6 mag fainter than the maximum brightness ever observed ($R=12.00$ on May 9, 2005). Indeed,
apart from the exceptional 2004--2005 outburst, the last-season outbursts are the most noticeable events that have been observed so far in the optical band for 3C 454.3 (see the top panel of Fig.\ \ref{storico}).

Besides the exceptional intranight variations in November--December 2007 reported by \citet{rai08b}, 
noticeable episodes of very fast variability were observed also during other outburst phases.
Figure \ref{idv} shows two enlargements of the $R$-band light curve. 
In the upper panel we can appreciate in detail the source activity in July 18--28, 2007. 
In the lower panel we can see the violent variations between December 31, 2007 and January 6, 2008.
The most remarkable episode, occurred on January 2 (JD = 2454467.5--2454468.5), includes a 0.55 mag brightness decrease in 1.37 hours, i.e.\ a rate of 0.0067 mag per minute, which is comparable to what observed on December 1, 2007 (JD = 2454435.5--2454436.5, see \citealt{rai08b}). These variations are among the most extreme flux changes detected in blazars.
They were observed when the source was in particularly bright states, which agrees with the idea that outbursts may be due to changes in the Doppler beaming factor. Indeed, an increase of this factor implies a magnification of the flux variations as well as a contraction of the variability time scales.

   \begin{figure}
      \caption{Enlargements of the $R$-band light curve in the periods July 18--28, 2007 (top) 
and December 31, 2007 -- January 6, 2008 (bottom).} 
         \label{idv}
   \end{figure}

\subsection{Colour analysis}

Colour analysis is an important tool to investigate the spectral behaviour of the source and, in turn, the nature of its emission.

In Fig.\ \ref{colori} we show the $R$-band light curve in the last four observing seasons (top panel) together with the corresponding $B-R$ colour indices as a function of time (middle panel) and of brightness level (bottom panel). 
The colour indices were calculated by selecting $B$ and $R$ data points with errors not greater than 0.10 and 0.05 mag, respectively, and by coupling $B$ and $R$ data taken by the same telescope within at most 20 minutes (typically $\sim 5$ minutes).
A total of 864 $B-R$ indices were obtained, whose average value is 1.11 mag, with a standard deviation of 0.13 mag.
   \begin{figure}
      \caption{The $R$-band light curve in the last four observing seasons (top panel) is shown
 together with the corresponding $B-R$ colour indices as a function of time (middle panel) and of 
brightness level (bottom panel). 
In the top panel the mean $R$-band mag (14.65) is shown as a black dotted line, and blue dots indicate those $R$ points that were used to calculate the $B-R$ indices shown in the panel below. In the middle panel the blue dashed line indicates the average $B-R$ colour index (1.11); the two red arrows highlight two bluer-when-brighter events. In the bottom panel the red line represents a parabolic fit to the data, while the green one represents the linear fit to the $F_B$ versus $F_R$ correlation.}
         \label{colori}
   \end{figure}

The behaviour of $B-R$ during the 2004--2005 outburst, up to JD $\sim$ 2453650, was analysed by \citet{vil06}. These authors noticed a general redder-when-brighter trend, which seems to ``saturate" for $R \la 14$. They ascribed this behaviour to the interplay of two emission components: a ``blue" contribution likely due to the thermal radiation from the accretion disc, and a ``red" contribution due to synchrotron radiation from the jet. The presence of the former would explain the redder-when-brighter behaviour seen in fainter states, while the latter would dominate the total emission in the brightest states, possibly imposing the bluer-when-brighter behaviour that has often been observed in blazars, especially of BL Lac type \citep[see e.g.][]{vil00,rai03,vil04a,pap07}. 
Taking into account the results obtained by \citet{rai07b}, we can say that the ``blue" emission component likely comes from the contribution of both the line emission from the broad line region (little blue bump) and the thermal emission from the accretion disc (big blue bump), which are expected to be less variable (and in any case on longer time scales) than the jet component.

The spectral behaviour during the exceptional 2004--2005 outburst can now be compared to the behaviour in the post-outburst phase and in the new activity period.
From Fig.\ \ref{colori} we can notice that also after JD $\sim$ 2453650 the source maintains its general redder-when-brighter behaviour, the lowest $B-R$ indices in particular corresponding to the 2006--2007 observing season, when the source was in a very faint state. 
In contrast, during the last-season outbursts, the ``saturation" effect appears.
This effect is also clearly visible in the $B-R$ versus $R$ plot, where a parabolic fit (red line) seems to better represent the data than a linear regression (whose coefficient is only $r=-0.784$).
The curvature is due to the fact that the synchrotron radiation for $R \la 14$ dominates the emission from the disc and broad line region.
The parabolic fit would furthermore imply a bluer-when-brighter trend for the brightest states. 

For comparison, we can follow the colour analysis approach adopted by \citet[][see also \citealt{hag94,hag07b}]{hag08}.
We plotted the observed flux density $F_B$ as a function of $F_R$, and found that the data points are best-fitted by a straight line $F_B=0.193+0.423 F_R$, with correlation coefficient $r=0.997$. 
A linear correlation of this kind is what we expect if the optical spectrum is due to an invariable ``blue" emission component plus a variable ``red" emission component with constant $B-R$.
If we report the result of the linear fit in the bottom panel of Fig.\ \ref{colori}
(green line) we see that it is very close to the parabolic fit.
The main difference is that instead of tracing a bluer-when-brighter trend in the brightest states, it asymptotically approaches the constant colour index of the variable jet component.
The paucity of $B-R$ indices corresponding to very bright states makes it difficult to distinguish between the two interpretations. 
However, we notice that in at least a couple of events belonging to the July--August 2007 outburst (highlighted by red arrows in the middle panel), we can recognise a bluer-when-brighter trend. 

\subsection{Millimetric and centimetric radio data}

Since the observation of the exceptional outburst in 2005, one of the major issues has been to investigate the correlation between the emission in the optical band and that at longer wavelengths. 
This was one of the main motivations that pushed the WEBT collaboration to continue the monitoring effort in the post-outburst phase. 
The picture that emerged was rather complex, and according to \citet{vil07} it can be explained if both geometrical effects and changes in the jet energetics are taken into account.
The new phase of activity that started in mid 2007 gives us the possibility to test this interpretation.

In Fig.\ \ref{radop} we show the optical flux density in the $R$ band\footnote{To convert magnitudes into de-reddened flux densities we corrected for Galactic extinction assuming $A_B$ = 0.462 from \citet{sch98} and deriving the values for the other bands according to \citet{car89}; zero-mag flux densities were taken from \citet{bes98}.} together with the radio light curves in the last four observing seasons.
The $R$-band light curve has been complemented with data points from other bands to minimize gaps due to solar conjunction. In particular, we added two data points (green diamonds) representing the minimum and maximum fluxes observed by Swift on April 24 and 25, 2005, during a flux increasing phase \citep[see also][]{gio06}. The data acquired with the UltraViolet and Optical Telescope (UVOT) instrument onboard Swift were reduced in the same way as in \citet{rai08b}; magnitudes in the $B$ band were converted to the $R$ band adopting a $B-R$ value of 1.2, which is adequate for bright states (see Fig.\ \ref{colori}).
These two data points highlight that the historical maximum observed on May 9, 2005 was part of a fast flare, belonging to a much longer outburst phase. Moreover, in the top panel of Fig.\ \ref{radop} we also included the $J$-band data acquired in May 2007 before the start of the optical observations (blue crosses). These were transformed to the $R$ band assuming an average colour index $R-J = 1.75$.

Radio data at both mm and cm wavelengths were collected as already calibrated flux densities.
Our light curves were complemented with data from the VLA/VLBA Polarization Calibration Database\footnote{\tt http://www.vla.nrao.edu/astro/calib/polar/}. 
   \begin{figure}
      \caption{$R$-band flux densities (mJy, top panel) in 2004--2008 compared to radio flux densities (Jy) 
at different frequencies. In the top panel, the two green diamonds indicate the minimum and maximum optical fluxes derived from Swift observations on April 24 and 25, 2005; blue crosses represent $J$-band data taken in May 2007 before the start of the optical observations (see the text for more details). Literature radio data from \citet{vil06,vil07} and \citet{rai07b} are displayed as black plus signs.
Vertical lines indicate the two XMM-Newton pointings of May 23 and 31, 2007. The solid black line in the second panel represents the predicted mm light curve according to the model by \citet{vil07}.
X-ray flux densities at 1 keV ($\mu$Jy, multiplied by a factor 2) were overplotted on the optical and mm light curves as orange plus signs.}
         \label{radop}
   \end{figure}

In the second panel we display the 1 mm light curve,  which is presented here for the first time.
After the 2005 solar conjunction, on May 6 (i.e.\ 3 days before the observation of the optical maximum brightness level) the source was observed with a flux density 2.7 times higher than in January.
Then the flux further increased, reaching $F_{\rm 1 \, mm}=42.40$ Jy on June 3. The source remained in a high mm state until early September. A minor mm bump was then observed in February 2006, which, according to \citet{vil07}, was connected to the minor optical flare of October--November 2005. The prediction of the \citet{vil07} model for the 1 mm light curve is plotted as a solid line; this result was obtained by ``interpolating" the trends observed in the optical and high-frequency radio bands.
The model reproduces fairly well the main trend of the light curve up to mid 2006.
Afterward the mm flux, like the optical one, remained low until the start of the last observing season.
Indeed, the July--August 2007 optical outburst had a mm broader counterpart; in particular,
the first peak at JD $\sim$ 2454300 seems to be contemporaneous in the two bands, while only a slight and delayed decrease of the mm flux appears to correspond to the optical outburst fall.

\begin{figure}
   \resizebox{\hsize}{!}{\includegraphics{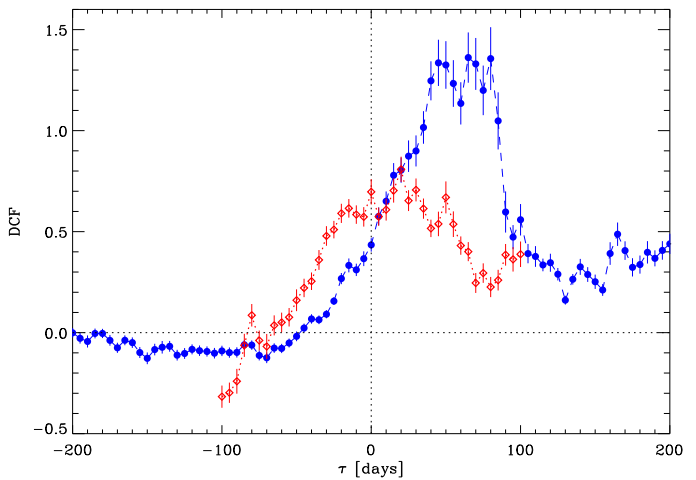}}
      \caption{Discrete correlation function (DCF) between the 1-day binned $R$-band and 1 mm light curves shown in Fig.\ \ref{radop} (blue filled circles); the result mainly reflects the optical and mm behaviour during the 2005 outburst. The red empty diamonds show the DCF obtained considering only the data after JD = 2454000.} 
         \label{dcf_rmm2}
   \end{figure}

To investigate the relationship between the optical and 1 mm flux variations, we performed a cross-correlation analysis by means of the discrete correlation function \citep[DCF; see][]{ede88,huf92,pet01}, a method that has been specifically conceived for handling unevenly-sampled data trains. A DCF peak (dip) means correlation (anti-correlation), which is stronger as the DCF value is greater. 
The position of the peak indicates the time lag between the flux variations in the two bands.
The DCF obtained by cross-correlating the $R$-band and 1 mm light curves shown in Fig.\ \ref{radop} is displayed  in Fig.\ \ref{dcf_rmm2} as blue filled circles. 
The broad peak reaching a high DCF value of $\sim 1.4$ implies that the correlation is strong, and that the mm flux changes lag behind the optical ones on a time scale ranging from about 40 to 80 days. The maximum of the signal is at 65 days, while the calculation of the centroid\footnote{The centroid is defined as $\tau_{\rm c}=\sum_i \tau_i {\rm DCF}_i / \sum_i {\rm DCF}_i$, where sums
run over the points which have a DCF value close to the peak one (typically not less than 70--80\%).} yields 60 days. This result is dominated by the behaviour during the 2005 outburst, when the sharp optical peak was soon followed by a broader mm maximum. 
Indeed, the DCF does not change significantly when considering only the data before JD = 2454000.
To check whether the correlation has changed in the recent activity phase, we calculated the DCF 
considering only outburst-free data, taken after JD = 2454000. The result is displayed as red empty diamonds in Fig.\ \ref{dcf_rmm2}.
The DCF now shows a broad and smaller, but still significant (DCF $\sim$ 0.8) maximum, indicating correlation with a mm delay peaking at 20 days and with centroid of $\sim 22$ days.

The shorter time delay of the mm flux variations with respect to the optical ones that characterises the last activity phase may suggest that the jet regions emitting the optical and mm radiations are now better aligned than in the past, and/or that the opacity in the jet has decreased, allowing the release of mm radiation closer to the optically emitting zone. 
This is what the observations allow us to infer. However, we must consider that both in 2005 and in 2007 the long observing gaps in the optical band due to solar conjunction may have hidden other flaring events. Consequently, the mm delay may have been longer.

The 43 GHz data shown in Fig.\ \ref{radop} present some scatter, but a cubic spline interpolation through the 30-day binned light curve allows us to follow the main trend at this frequency. 
The maximum value in the last observing season is reached at JD $\sim$ 2454420.
The DCF between the $R$-band and 43 GHz data in the last observing season presents a peak at 110 days. This delay of the 43 GHz variations with respect to the optical ones is much shorter than the 275 day lag between the 2005 optical outburst and the peak at 43 GHz in early 2006. But it is comparable to the 125 day time delay estimated by \citet{vil07} assuming that the 43 GHz counterpart of the big optical outburst is the broad hump peaking at JD $\sim$ 2453640, while the stronger and faster radio event peaking at JD $\sim$ 2453800 is actually connected to the optical flare occurred in October--November 2005. According to their interpretation, starting from May 2005 the optically emitting region became less aligned with the line of sight, with a consequent decrease of the Doppler beaming factor. In contrast, the viewing angle of the radio emitting region decreased, leading to an increase of the Doppler factor. This would explain why the big optical outburst of 2004--2005 produced a smaller radio event than that produced by the modest October--November 2005 optical flare.

The behaviour of the 37 GHz light curve is very similar to that at 43 GHz. Cross-correlation between the optical and 37 GHz data in the last observing season yields a radio lag of 130 days.
The DCF value is lower than in the optical-43 GHz case ($\sim 0.5$ instead of $\sim 0.8$)
because the 37 GHz light curve is smoother than the 43 GHz one.
The 43-37 GHz cross-correlation in the last season presents a very broad peak whose maximum corresponds to a lag of 20 days, but whose centroid is only 2--6 days. This is in agreement with the 5 day time lag found by \citet{vil07} for the 2005 outburst.
The variations observed in the 43 and 37 GHz light curves are less and less prominent going toward lower frequencies, disappearing completely at 5 GHz. This is likely due to the strong opacity of the emitting regions at these frequencies. 

\subsection{Historical radio-to-optical emission behaviour}

The historical optical and radio light curves of 3C 454.3 are shown in Fig.\ \ref{storico}.

The data taken by the WEBT in the last observing seasons highlight a decreasing trend and lack of variability of the lower-frequency radio flux. Indeed, the last outburst that has been observed below $\sim 15 $ GHz dates back to 1994.
This suggests that the jet region producing the low-frequency radio emission has become
progressively less aligned with the line of sight \citep{vil06}.

In contrast, the high activity phase in the optical band that started in 2001 could mean that the optical emitting region is seen under a smaller viewing angle than it was in the past, so that flux changes caused e.g.\ by the passage of a disturbance are enhanced by an increased Doppler factor. 

The occurrence of changes in the jet orientation is supported by the analysis of VLBI maps at different epochs: indeed, \citet{jor05} found a displacement of the VLBI core between 1997.6 and 1998.2 that is expected to cause a change in the viewing angle of the innermost jet.

The higher-frequency radio emission (above 8 GHz) shows a behaviour that seems the result of two contributions. One is a ``hard" component, originating close to the optically emitting region and connected to the optical emission. The other one is a ``soft" component coming from an outer and more transparent jet region, which carries the signature of the events observed at longer wavelengths.
We notice that the existence of two different kinds of radio events in blazars has long been recognised \citep[e.g.][]{val92,rai03,vil04b,bac06,rai08a}

   \begin{figure}
      \caption{Historical emission behaviour of 3C 454.3 in the optical band and in various radio bands. 
In the top panel, the $R$-band light curve (red dots) starting from 1990 has been complemented with
historical $B$-band data (blue circles) before that date, shifted by a colour index $B-R=1.0$, which is appropriate for intermediate--faint states (see Fig.\ \ref{colori}).} 
         \label{storico}
   \end{figure}

\section{Near-IR spectra}

   \begin{figure}
   \resizebox{\hsize}{!}{\includegraphics[width=8cm,clip]{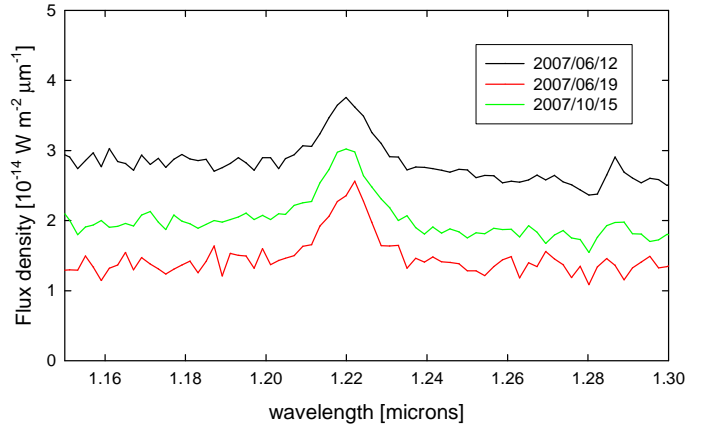}}
      \caption{Three out of the sixteen near-IR spectra acquired with the 1.1 m telescope at Campo Imperatore.
They show a prominent H$\alpha$ emission line at 1.22 $\mu \rm m$.}
         \label{spectra}
   \end{figure}

There are relatively few studies investigating the relationship between the continuum and line fluxes in blazars \citep{cor00,rai07a}.
The reason is that lines may be difficult to measure in these sources, especially during bright states.
We performed a spectroscopic monitoring of 3C 454.3 in the near-IR, where a pronounced H$\alpha$ emission line was expected \citep{rai07b}, because of the source redshift $z=0.859$.
The spectra were acquired in June--July and October--November 2007, in periods when the source was not in a flaring state. 

All spectra were obtained at the 1.1 m AZT-24 telescope located at Campo
Imperatore (L'Aquila, Italy), equipped with the imager/spectrometer SWIRCAM \citep{dal00}. 
Low-resolution ($R \approx 250$) spectroscopy was obtained by means of an IR grism
covering the $ZJ$ (0.83--$1.34 \, \mu \rm m$) band, taking subsequent exposures
with a total integration time of $\approx 1200$ s.
The observations were flat-fielded and sky-subtracted.
Telluric and instrumental features were removed by dividing the extracted
spectra by that of a normalised telluric standard star, once corrected for its
intrinsic spectral features. Flux calibration was obtained from our photometric
data. 
After subtraction of continuum using the neighbouring part of the spectrum, the H$\alpha$ emission profile was fitted by a Gaussian function. The parameters of this Gaussian (FWHM and peak intensity) were used to calculate the equivalent width (EW) and total flux radiated in line.

A selection of the sixteen spectra is displayed in Fig. \ref{spectra}.
In all the spectra a prominent H$\alpha$ emission line is visible.
The line is at 1.22 $\mu \rm \rm m$, which is the $\lambda_{\rm eff}$ of the $J$ filter, and this explains the flux excess in this band shown by the SEDs presented by \cite{rai07b}.
The observed EW ranges between 50 and 120 \AA.
The results of the spectral monitoring are shown in detail in Table \ref{line}, where 
Col.\ 1 reports the date of observation, 
Col.\ 2 the magnitude in $J$ band, 
Col.\ 3 the $J$ magnitude after correction for the H$\alpha$ line contribution, 
Col.\ 4 the line EW, 
Col.\ 5 the line flux, 
Col.\ 6 the continuum flux density, 
Col.\ 7 the line FWHM, and
Col.\ 8 the H$\alpha$ peak intensity relative to continuum.

\begin{figure}
   \resizebox{\hsize}{!}{\includegraphics{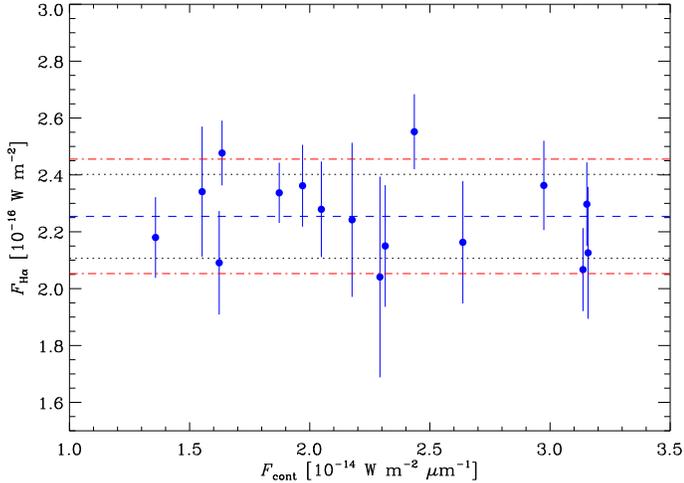}}
      \caption{H$\alpha$ emission line flux as a function of continuum flux density.
The average value of the line flux is marked with a blue dashed line; 
1 $\sigma$ deviations are indicated by black dotted lines, while deviations by 1 root mean square uncertainty are shown by the red dash-dotted lines.}
         \label{linecont}
   \end{figure}

The behaviour of the line flux as a function of the continuum flux density is displayed in Fig.\ \ref{linecont}.
We notice a factor 2.33 variation in the continuum flux density, while the change in the H$\alpha$ flux is only a factor 1.25, and no correlation is recognisable between the two quantities.
The average line flux is $2.25 \times 10^{-16} \rm \, W \, m^{-2}$, with a standard deviation of $0.15 \times 10^{-16} \rm \, W \, m^{-2}$ that is lower than the root mean square uncertainty 
$\epsilon_{\rm rms}=\sqrt{\sum_{i=1}^{N}{\epsilon_i ^2}/N}=0.20 \times 10^{-16} \rm \, W \, m^{-2}$ (where $\epsilon_i$ are the individual errors).
This means that the H$\alpha$ line flux variation is not statistically significant; hence the line emission seems not to be affected by the continuum variability.
This result agrees with the scenario according to which the blazar continuum is dominated by the beamed and variable jet emission, while the line is produced in the broad line region that is likely photoionised by the thermal radiation coming from the accretion disc.

\begin{table*}
\caption{Results of the near-IR spectroscopic monitoring at Campo Imperatore in 2007.}
\label{line}
\centering
\begin{tabular}{l c c c c c c c  }
\hline\hline
Date    & $J$ & $J_{\rm corr}$ & $\rm EW_{H \alpha}$ & $F_{\rm H \alpha}$ & $F_{\rm cont}$ & FWHM & Ratio\\
        &[mag]& [mag] & [\AA] &[$10^{-16} \rm \, W \, m^{-2}$]&[$10^{-14}\rm \, W \, m^{-2} \, \mu m^{-1}$]&[\AA]& \\
\hline
2007 Jun 12 & 12.804 $\pm$  0.014 &  12.847 &  79.3 &  2.552 $\pm$  0.132 &  2.435 &  150 &  0.174\\
2007 Jun 19 & 13.415 $\pm$  0.013 &  13.474 &  117.9 &  2.180 $\pm$  0.142 &  1.358 &  130 &  0.308\\
2007 Jun 22 & 13.327 $\pm$  0.017 &  13.389 &  118.3 &  2.341 $\pm$  0.229 &  1.552 &  170 &  0.266\\
2007 Jun 27 & 12.508 $\pm$  0.013 &  12.535 &  50.0 &  2.126 $\pm$  0.232 &  3.158 &  160 &  0.115\\
2007 Jun 29 & 12.706 $\pm$  0.006 &  12.742 &  62.0 &  2.163 $\pm$  0.215 &  2.637 &  140 &  0.161\\
2007 Jul 01 & 12.511 $\pm$  0.013 &  12.546 &  57.7 &  2.297 $\pm$  0.147 &  3.153 &  140 &  0.114\\
2007 Jul 02 & 12.525 $\pm$  0.009 &  12.555 &  51.3 &  2.067 $\pm$  0.146 &  3.137 &  150 &  0.117\\
2007 Oct 08 & 12.583 $\pm$  0.018 &  12.619 &  60.5 &  2.363 $\pm$  0.157 &  2.974 &  150 &  0.125\\
2007 Oct 11 & 12.941 $\pm$  0.029 &  12.987 &  82.1 &  2.242 $\pm$  0.271 &  2.176 &  150 &  0.197\\
2007 Oct 14 & 13.240 $\pm$  0.033 &  13.301 &  101.4 &  2.091 $\pm$  0.182 &  1.623 &  130 &  0.227\\
2007 Oct 15 & 13.089 $\pm$  0.036 &  13.140 &  95.4 &  2.337 $\pm$  0.106 &  1.873 &  150 &  0.218\\
2007 Oct 16 & 13.240 $\pm$  0.040 &  13.295 &  111.5 &  2.477 $\pm$  0.114 &  1.635 &  130 &  0.292\\
2007 Oct 26 & 12.892 $\pm$  0.041 &  12.933 &  71.1 &  2.041 $\pm$  0.353 &  2.292 &  110 &  0.226\\
2007 Oct 27 & 13.036 $\pm$  0.038 &  13.090 &  94.6 &  2.362 $\pm$  0.144 &  1.970 &  120 &  0.231\\
2007 Oct 28 & 12.865 $\pm$  0.023 &  12.908 &  71.3 &  2.150 $\pm$  0.214 &  2.314 &  120 &  0.205\\
2007 Nov 03 & 13.005 $\pm$  0.055 &  13.052 &  83.4 &  2.279 $\pm$  0.168 &  2.048 &  160 &  0.197\\
\hline
\end{tabular}
\end{table*}

\section{XMM-Newton observations}

As part of a long-term WEBT project to study the multifrequency variability of 3C 454.3 after the 2004--2005 outburst, 
the X-ray Multi-Mirror Mission (XMM) - Newton satellite observed the source in July 2--3 and December 18--19, 2006, and then on May 23 and 31, 2007 (PI: C.~M. Raiteri). The results of the first two observations were reported in \citet{rai07b}. 
On May 23, 2007 most of the observation was lost because of telecommanding problems, 
and re-scheduled for May 31.
However, the May 23 exposure was long enough to make data reduction worthwhile.

\subsection{EPIC data}

\begin{figure}
   \resizebox{\hsize}{!}{\includegraphics[angle=-90]{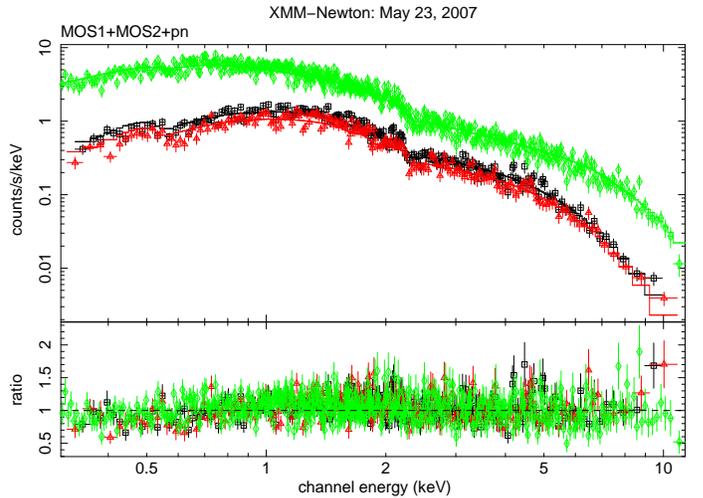}}
      \caption{EPIC spectrum of 3C 454.3 on May 23, 2007; 
black squares, red triangles, and green diamonds represent MOS1, MOS2, and pn data, respectively.
The bottom panel shows the ratio between the data and the folded model, a power law with Galactic absorption.} 
         \label{xmm3}
   \end{figure}

\begin{figure}
   \resizebox{\hsize}{!}{\includegraphics[angle=-90]{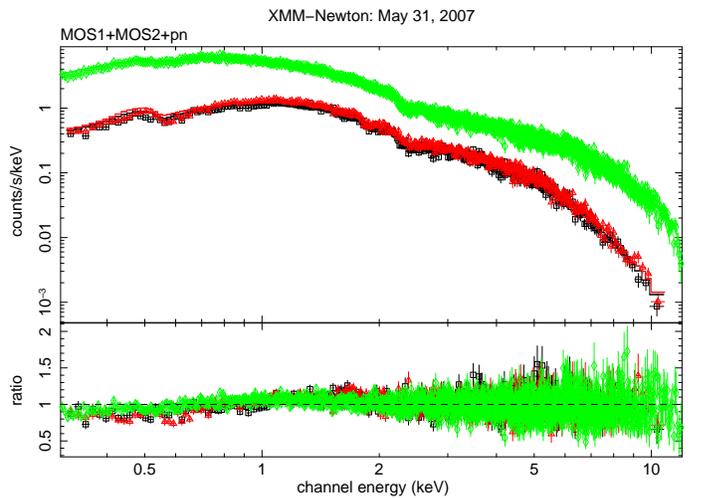}}
      \caption{EPIC spectrum of 3C 454.3 on May 31, 2007; 
black squares, red triangles, and green diamonds represent MOS1, MOS2, and pn data, respectively.
The bottom panel shows the ratio between the data and the folded model, a power law with Galactic absorption.} 
         \label{xmm4}
   \end{figure}

The European Photon Imaging Camera (EPIC) onboard XMM-Newton includes three detectors: 
MOS1, MOS2 \citep{tur01}, and pn \citep{str01}.
Data were reduced with the Science Analysis System (SAS) software, version 7.1,
following the same standard procedure adopted in \cite{rai07b}. The final analysis was performed with the {\tt Xspec} task of the XANADU package.
We fitted the MOS1, MOS2, and pn spectra of each epoch together to increase the statistics.
We first considered a single power-law model with Galactic absorption according to the \citet{wil00}
prescriptions, and $N_{\rm H}=0.724 \times 10^{21} \, \rm cm^{-2}$, from the Leiden/Argentine/Bonn (LAB) Survey \citep[see][]{kal05}.
The results of this spectral fitting are displayed in the top panels of 
Fig.~\ref{xmm3} (May 23) and Fig.~\ref{xmm4} (May 31); 
the bottom panels show the ratio between the data and the folded model. 
The corresponding model parameters are reported in Table~\ref{pow}, where 
Col.~2 gives the column density,
Col.~3 the photon spectral index $\Gamma$,
Col.~4 the unabsorbed flux density at 1 keV,
Col.~5 the 2--10 keV observed flux,
and Col.~6 the value of $\chi^2/\nu$ (with the number of degrees of freedom $\nu$).
The high $\chi^2/\nu$ values indicate that the folded model (single power law with Galactic absorption)
does not fit the data very well,
especially in the May 31 case.
The plots suggest that absorption in the soft 
X-rays has been underestimated by the model.

\begin{table*}
\caption{Results of fitting the EPIC data of May 23 and 31, 2007 with different models}             
\label{pow}      
\centering  
\begin{tabular}{ c c c c c c}  
\hline\hline            
Date & $N_{\rm H}$ & $\Gamma$ & $F_{\rm 1 \, keV}$ & $F_{\rm 2-10 \, keV}$  & $\chi^2/\nu$ ($\nu$) \\
     & [$10^{21} \, \rm cm^{-2}$]&   & [$\mu$Jy]       & [$\rm erg \, cm^{-2} \, s^{-1}$]  &   \\     
\hline                         
\multicolumn{6}{c}{Single power law with Galactic absorption}\\
2007 May 23  & 0.724 & 1.49  $\pm$ 0.01  & 3.39 $\pm$ 0.04 & $2.89 \times 10^{-11}$ & 1.19 (947)\\      
2007 May 31  & 0.724 & 1.532 $\pm$ 0.004 & 3.41 $\pm$ 0.01 & $2.73 \times 10^{-11}$ & 1.47 (2240)\\
\hline                                   
\multicolumn{6}{c}{Single power law with free absorption}\\
2007 May 23  & 1.12 $\pm$ 0.07 & 1.60 $\pm$ 0.02   & 3.84 $\pm$ 0.09 & $2.75 \times 10^{-11}$ & 1.07 (946)\\      
2007 May 31  & 1.09 $\pm$ 0.02 & 1.627 $\pm$ 0.007 & 3.80 $\pm$ 0.03 & $2.61 \times 10^{-11}$ & 1.11 (2239)\\
\hline
\multicolumn{6}{c}{Double power law with fixed extra absorption}\\
2007 May 23  & 1.34  & 1.55, 2.61 $^{+0.47}_{-0.42}$ & 3.99 $^{+0.13}_{-0.23}$ & $2.77 \times 10^{-11}$ & 1.07 (946)\\      
2007 May 31  & 1.34  & 1.578, 2.84 $\pm$ 0.13 & 3.98 $^{+0.04}_{-0.05}$ & $2.63 \times 10^{-11}$ & 1.13 (2239)\\
\hline
\end{tabular}
\end{table*}

The need of extra absorption to fit the X-ray spectra of this source has already been pointed out by several authors \citep[see e.g.][for a review]{rai07b}.
Indeed, when fitting a power-law model with free absorption, the goodness of fit improves (see Table \ref{pow}).
The best-fit $N_{\rm H}$ values are about  $1.1 \times 10^{21} \, \rm cm^{-2}$, i.e.\ 52\% higher than the Galactic amount,
and 26\% and 9\% higher than those determined by \citet{rai07b} when fitting the July 2--3 and December 18--19, 2006 data, respectively.
At the same time, the 1 keV flux density has increased by a factor $\sim 4.2$ with respect to July and by $\sim 2.9$ with respect to December. 

This increase of the $N_{\rm H}$ value with source brightness would be difficult to explain in terms of variability of the column density; more likely it indicates that the spectral shape changes with flux.
\citet{rai07b} discussed the possibility that the X-ray spectrum presents a curvature.
They adopted a double power-law model with absorption fixed at $N_{\rm H}=1.34 \times 10^{21} \, \rm cm^{-2}$, as obtained by \citet{vil06}\footnote{This value was misprinted in that article: 13.40 instead of 1.34.} 
from the analysis of the Chandra observation of May 2005, during the outburst phase.
To reduce the uncertainties, \citet{rai07b} also fixed one of the two spectral indices by fitting the data 
above 2 keV with a single power law, since absorption plays a negligible role at those energies.
We investigated this possibility, following the same procedure; the results are shown in Table \ref{pow}.
From a statistical point of view, this last spectral fit is more or less as good as the previous one, but it provides 
an alternative view to the  $N_{\rm H}$-flux correlation, replacing it with a spectral softening-flux anticorrelation, 
according to which the low-energy part of the X-ray spectrum becomes softer as the flux decreases. 

\begin{figure}
   \resizebox{\hsize}{!}{\includegraphics{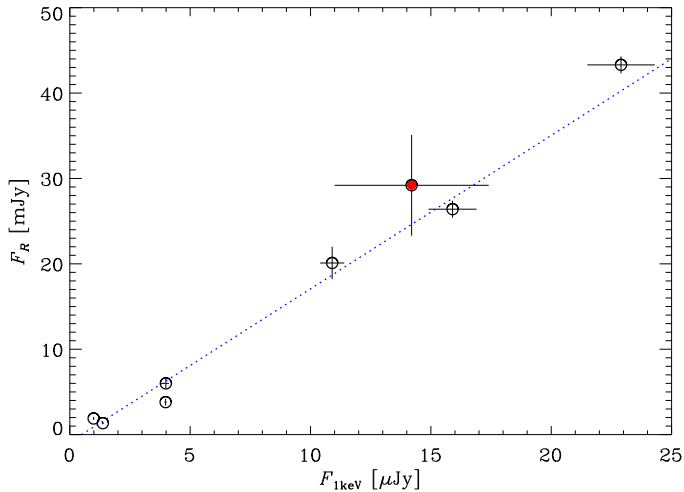}}
      \caption{$R$-band versus 1 keV flux density. The vertical error bars take into account 
the optical variations during the X-ray observations, which were particularly dramatic on May 19, 2005 (red filled circle).}
         \label{ox}
   \end{figure}

Finally, we investigate the connection between the X-ray flux and the emission at lower frequencies.
The X-ray radiation is commonly believed to be produced by inverse-Compton scattering of low-energy photons off relativistic electrons in the plasma jet. Consequently, we expect that the X-ray fluxes correlate with the synchrotron emission at some frequency.
In the first two panels of Fig.\ \ref{radop} we plotted the 1 keV flux densities derived from Chandra \citep{vil06}, Swift \citep{gio06}, and XMM-Newton \citep[][and this work]{rai07b} observations. The X-ray data match both the optical and mm data fairly well in faint states, but during the 2005 outburst there seems to be a good agreement with the optical light curve but not with the mm one, unless we missed some very rapid flare in this band. However, the latter hypothesis appears rather unlikely, since events of this kind are not observed at other epochs. 
The correlation between the optical and X-ray flux densities is shown in Fig.\ \ref{ox}. The vertical error bars take into account the 
optical variability during the X-ray observations. This was particularly violent on May 19, 2005, when both Swift and Chandra were observing, so that also the X-ray data were averaged. The correlation coefficient is $r=0.992$.

This result suggests that the seed photons for the inverse-Compton process producing the X-ray radiation may be synchrotron photons at IR--optical frequencies, possibly upscattered by colder electrons. Alternatively, whatever the photons are, the relativistic electrons upscattering them are the same that produce the IR--optical synchrotron emission. Hence, the production site of the X-ray radiation is likely that of the IR--optical radiation. This region is located more internally in the jet than that producing the mm emission.

\subsection{OM data}

The XMM-Newton satellite also carries a co-aligned 30 cm
optical--UV telescope \citep{mas01}, the Optical Monitor (OM).
Because of the short duration of the May 23 observation we were left with only one exposure in the UV$M2$ band, while on May 31 observations in $B$, $U$, UV$W1$, and UV$M2$ bands were performed.
The OM data were reduced with the {\tt omichain} task of SAS version 7.1 and analysed with
{\tt omsource}. The results are reported in Table~\ref{om}, where
Cols.~2 and 5 display the exposure times for each filter for the May 23 and May 31 observations, respectively;
Cols.~3 and 6 the source magnitudes, and 
Cols.~4 and 7 the de-reddened source flux densities in mJy.
The latter were obtained by following the same prescriptions used by \citet{rai07b} for the Galactic extinction and flux conversion through Vega.

The errors on the OM photometry are quite small and may appear underestimated; however, we must consider the long exposure times. Indeed, a confirmation of the good quality of the OM data in the optical bands comes from the comparison with the ground-based photometry of the reference stars; if we consider stars of about the same brightness as the source, the OM photometry is in agreement with the \citet{ang71} and the \citet{rai98} photometry adopted in this work within a few hundredths of magnitude.
In the UV, where this comparison is not possible, we notice that the six subsequent exposures in the UV$M2$ band of May 31 yield a mean value of 15.34 mag for the source, with a standard deviation of 0.03 mag only.

\begin{table*}
\caption{Results of the OM observations of 3C 454.3 on May 23 and 31, 2007.}
\label{om}      
\centering  
\begin{tabular}{ c | c c c | c c c}  
\hline\hline            
       & \multicolumn{3}{|c|}{2007 May 23}        & \multicolumn{3}{|c}{2007 May 31}\\
Filter & $t_{\rm exp}$ & Brightness & $F_\nu$ & $t_{\rm exp}$ & Brightness & $F_\nu$\\
       & [s] &  [mag]            & [mJy]             & [s]  & [mag]            & [mJy]\\     
\hline                        
$B$    & -    & -                & -                 & 1400 & 16.20 $\pm$ 0.01 & 2.134 $\pm$ 0.020\\      
$U$    & -    & -                & -                 & 1400 & 15.44 $\pm$ 0.01 & 1.603 $\pm$ 0.014\\
$W1$   & -    & -                & -                 & 3000 & 15.26 $\pm$ 0.01 & 1.521 $\pm$ 0.014\\
$M2$   & 2679 & 15.05 $\pm$ 0.03 & 1.857 $\pm$ 0.051 &18758$^a$ & 15.34 $\pm$ 0.03$^b$ & 1.422 $\pm$ 0.039\\
\hline                                   
\multicolumn{7}{l}{$^a$ Six subsequent exposures of 3120, 3119. 3121, 3120, 3119, and 3159 s.}\\
\multicolumn{7}{l}{$^b$ Average over the results of the six subsequent exposures.}
\end{tabular}
\end{table*}

\section{Spectral energy distributions}

\begin{figure*}
    \sidecaption
    \includegraphics[width=12cm]{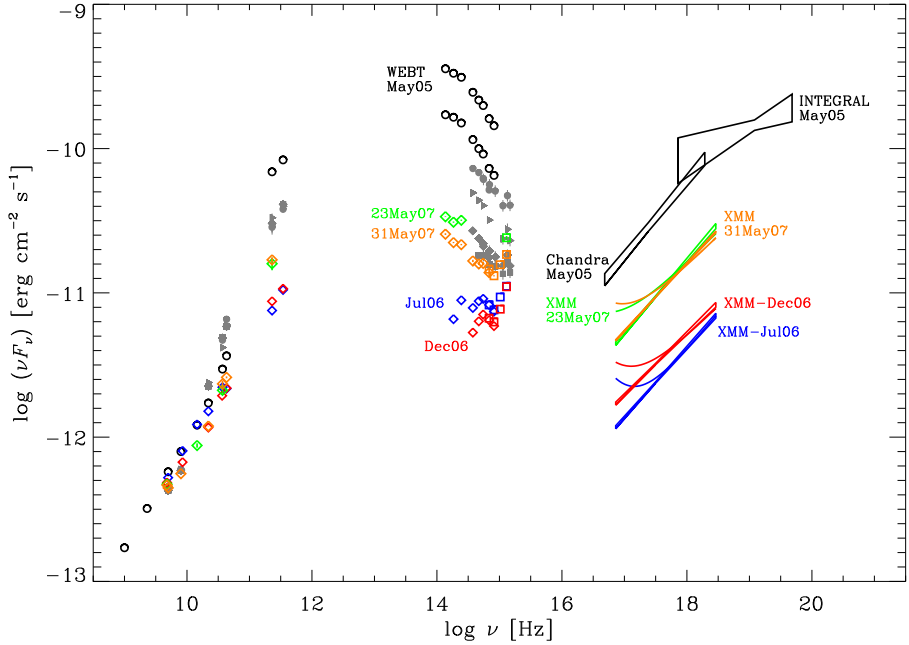}
      \caption{Broad-band SED of 3C 454.3 at various epochs. Radio-to-optical data acquired by the WEBT in May 2005 are displayed as black circles; in the optical they indicate the range of variability observed during the Chandra and INTEGRAL pointings. Grey filled symbols refer to the SEDs presented by \citet{rai08b}, built with WEBT and Swift-UVOT data taken in November--December 2007. Coloured SEDs show data obtained during observations by XMM-Newton; the July and December 2006 data are from \citet{rai07b}. Both a power law and a double power law fits to the EPIC spectra are reported (see text for details); the OM data are indicated by empty squares and the WEBT data by empty diamonds.}
         \label{sed}
   \end{figure*}

In Fig.\ \ref{sed} we show the broad-band SED of 3C 454.3 at various epochs.
We report the high brightness state observed in May 2005 by the WEBT (radio-to-optical) and by Chandra and INTEGRAL \citep{vil06,pia06}, as well as the two SEDs corresponding to the XMM-Newton pointings of July and December 2006 \citep{rai07b}. 
The above SEDs are now complemented by mm data from this work.
Moreover, we show the SEDs obtained with WEBT plus Swift-UVOT data acquired in November--December 2007 that were presented by \citet{rai08b}.
Finally, we display the data referring to the two XMM-Newton observations of May 23 and 31, 2007. 
Both the power-law with Galactic absorption and the double power-law spectral fits to the EPIC data (see Sect.\ 4.1) are shown.
The X-ray spectra acquired on May 23 and 31 show roughly the same slope as the spectra taken by XMM-Newton in July and December 2006, and by Chandra in May 2005, and an intermediate flux level.
However, the double power-law fits indicate a slighter curvature in the soft X-ray frequency range than in 2006. This curvature would allow a reconnection with the UV part of the SED, and the fact that the curvature is stronger when the flux is lower corresponds to the relatively more prominent UV excess in fainter states. 

The optical--UV flux in May 2007 was higher than in 2006, and the little blue bump observed by \citet{rai07b} is hardly recognisable. In 2007 this bump, due to line emission from the broad line region, is likely overwhelmed by the beamed synchrotron radiation from the jet. The jet emission however is not strong enough to completely hide the UV excess that \citet{rai07b} ascribed to thermal emission from the accretion disc (big blue bump). 
Similarly, in the May 23 and 31 near-IR data a small excess in the $J$ band is still visible, as the signature of the H$\alpha$ emission line discussed in Sect.\ 3.

The Swift-UVOT data presented by \citet{rai08b} suggest that the UV excess may be less prominent than what is shown by the OM data.
This may depend at least in part on the difference in the $\lambda_{\rm eff}$ of the UV filters in the two instruments. Indeed, the above authors have already stressed the large uncertainty affecting the UV data deriving from the correction for Galactic extinction.
This correction strongly depends on the considered wavelength in the UV, and it is as large as $\sim 1$ mag in the UV$M2$ and UV$W2$ bands.
However, the main point is that both the UVOT and OM data indicate the same trend, i.e.\ a larger UV excess for decreasing flux.

Noticeable variability is also visible at mm wavelengths.
The 1 mm datum in the May 23 SED has been obtained by averaging the data of May 18 and 29; that in the May 31 SED by averaging the May 29 and June 1 data. The November--December 2007 mm data indicate much brighter states than in May 2007 (see also Fig.\ \ref{radop}), even when the corresponding optical--UV data are approximately at the same level.

Radio data at cm wavelengths were acquired within 5 days before/after the XMM-Newton pointings of May 2007. 
For the high radio frequencies (43, 37, and 22 GHz), the behaviour is similar to that in the mm range.
At the lower radio frequencies, flux changes between May and November--December 2007 are less and less significant, as the flatness of the corresponding light curves in Fig.\ \ref{radop} confirms.

\section{Conclusions}

In the 2007--2008 observing season, the blazar 3C 454.3 underwent a new activity phase, which was monitored from the radio to the optical bands in the ambit of a WEBT campaign.
XMM-Newton observations in May 2007 and near-IR spectroscopic monitoring complemented the observing effort.
These new data traced the recent behaviour of the source that was compared to the old one, especially during the 2005 exceptional outburst. In particular, we presented for the first time the historical mm light curve, which fills the gap between the radio and optical frequency ranges. In the following we summarise the main results of this paper.

\begin{itemize}
\item We detected an optical-mm outburst in July--August 2007, and several faster optical events in November 2007 -- February 2008. 
During these bright phases we observed several episodes of intranight variability. The fact that fast variability can only be seen in flaring states may indicate that it is due to changes in the Doppler factor.

\item Outbursts were observed also at the higher radio frequencies (43--37 GHz), the flux enhancement decreasing at the lower radio frequencies, and disappearing below $\sim 15$ GHz.
The delays with which the radio outbursts followed the July--August 2007 optical outburst (3--4 months at 43--37 GHz) are similar to those claimed by \citet{vil07} for the 2005--2006 events. Moreover, we found that the prediction of \citet{vil07} for the mm light curve fairly matches the mm data. These results support their interpretation, suggesting that besides intrinsic processes, variations in the jet orientation are likely to play an important role in the source flux variability. They may also account for the change in the emission behaviour after $\sim 2000$, when outbursts appeared in the optical light curves and disappeared at the longest radio wavelengths (below 15 GHz).
The occurrence of changes in the jet orientation of this source is confirmed by the analysis of VLBI maps at different epochs \citep{jor05}.

\item The behaviour of the optical spectrum in the last two observing seasons followed the same patterns as during the 2004--2005 exceptional outburst.
The variation in the $B-R$ index as a function of brightness is not linear, but it is well described by a parabolic fit. Indeed, the optical spectrum of the source is affected by various emission components (synchrotron radiation, little and big blue bumps), with their own colours. A redder-when brighter trend characterises the fainter states, until $R \sim 14$, when this trend ``saturates" due to the fact that the jet component becomes dominant \cite[see also][]{vil06}. This jet component likely has an intrinsically variable spectrum, which follows a bluer-when-brighter trend.

\item The near-IR spectroscopic monitoring reveals the presence of a prominent H$\alpha$ broad emission line, thus explaining the flux excess in the $J$ band clearly visible in faint state SEDs. The line flux does not vary significantly as the continuum flux changes, suggesting that also in this source the broad line region is not photoionised by the variable jet emission, but likely by the thermal emission from the accretion disc.

\item Our analysis of new data from XMM-Newton, compared to previous observations, allows us to recognise a correlation between the UV excess and the excess in the soft X-rays. This confirms the presence of that thermal component from the accretion disc that is considered responsible for the H$\alpha$ broad emission line in the near-IR spectra.

\item We constructed a 1 keV light curve with data from the literature and from this work, and compared it to the light curves at lower frequencies. The X-ray flux correlates with the optical one.
This suggests that either the seed photons for the inverse-Compton process producing the X-ray radiation are synchrotron photons at IR--optical frequencies (possibly upscattered by colder electrons) or that, no matter the origin of the photons, the relativistic electrons upscattering them are the same producing the optical synchrotron emission. The X-ray radiation is thus likely produced in the jet region where the IR--optical emission comes from. 

\end{itemize}

The GLAST-AGILE Support Program of the WEBT \citep[see][]{vil08} is continuing the monitoring of 3C 454.3 in the just-started 2008--2009 observing season. New activity has already been observed at all frequencies \citep{vil08atel}, and the AGILE and GLAST satellites have detected strong $\gamma$-ray events \citep{don08atel,vit08atel,gas08atel,tos08atel,pit08atel}.

\begin{acknowledgements}
This work is partly based on observations made with the Nordic Optical Telescope, operated
on the island of La Palma jointly by Denmark, Finland, Iceland,
Norway, and Sweden, in the Spanish Observatorio del Roque de los
Muchachos of the Instituto de Astrofisica de Canarias, and on observations collected at the German-Spanish Calar Alto Observatory, jointly operated by the MPIA and the IAA-CSIC.
AZT-24 observations are made within an agreement between  Pulkovo, Rome and Teramo observatories.
The Submillimeter Array is a joint project between the Smithsonian Astrophysical Observatory and the Academia Sinica Institute of Astronomy and Astrophysics and is funded by the Smithsonian Institution and the Academia Sinica.
This research has made use of data from the University of Michigan Radio Astronomy Observatory,
which is supported by the National Science Foundation and by funds from the University of Michigan.
The Mets\"ahovi team acknowledges the support from the Academy of Finland.
This work is partly based on observation from Medicina and Noto telescopes operated by INAF - Istituto di Radioastronomia and the 100-m telescope of the MPIfR (Max-Planck-Institut für Radioastronomie) at Effelsberg.
The Torino team acknowledges financial support by the Italian Space Agency through contract 
ASI-INAF I/088/06/0 for the Study of High-Energy Astrophysics.
Acquisition of the MAPCAT data at the Calar Alto Observatory is supported in part by the Spanish ``Ministerio de Ciencia e Innovaci\'on" through grant AYA2007-67626-C03-03.
This paper is partly based on observations carried out at the IRAM 30-m telescope. IRAM is supported by INSU/CNRS (France), MPG (Germany) and IGN (Spain). 
\end{acknowledgements}

\end{document}